\documentclass[aps,prl,twocolumn,superscriptaddress,showpacs,amsmath,amssymb,longbibliography]{revtex4-1}
\usepackage{graphicx}
\usepackage[breaklinks]{hyperref}

\usepackage[utf8]{inputenc}
\usepackage{polski}
\usepackage[english]{babel}
\usepackage{amsmath}
\usepackage{xcolor}
\usepackage{ulem}

\definecolor{nred} {RGB}{224,0,0}

\definecolor{nblue} {RGB}{28,130,185}

\definecolor{nred} {RGB}{224,0,0}

\makeatletter
\newcommand*{\balancecolsandclearpage}{%
  \close@column@grid
  \clearpage
  \twocolumngrid
}

\usepackage[sort&compress]{natbib}
 
\begin{document}

\title{Spin subdiffusion in  disordered Hubbard chain}%
\author {Maciej Kozarzewski}
\affiliation{Institute of Physics, University of Silesia, 40-007 Katowice, Poland}
\author{Peter Prelov\v sek}
\affiliation{J. Stefan Institute, SI-1000 Ljubljana, Slovenia }
\affiliation{Faculty of Mathematics and Physics, University of Ljubljana, SI-1000 Ljubljana, Slovenia }
\author{Marcin Mierzejewski}
\affiliation{Department of Theoretical Physics, Faculty of Fundamental Problems of Technology, Wroc\l aw University of Science and Technology, 50-370 Wroc\l aw, Poland}

\begin{abstract}
We derive and study the effective spin model that explains the anomalous spin dynamics in the one-dimensional Hubbard model with strong potential disorder. 
Assuming that charges are localized, we show that spins are delocalized and their subdiffusive transport originates from a singular 
random distribution of spin exchange interactions.  The exponent relevant for the subdiffusion  is determined by the Anderson localization length 
and the density of electrons. While the analytical derivations are valid for low particle density, numerical results for the full model reveal a qualitative
agreement up to half--filling.
\end{abstract}

\maketitle

{\it Introduction.--}
The many--body localization  (MBL) \cite{Basko06,Oganesyan07} has recently  been intensively studied.  Vast amount of numerical data allowed to identify the main properties of the MBL systems:
 vanishing steady transport, \cite{berkelbach10,barisic10,agarwal15,lev15,steinigeweg15,barisic16,kozarzewski16,prelovsek217}
 absence of thermalization \cite{pal10, Serbyn2013, lev14,schreiber15,serbyn15,khemani15,luitz16,gornyi05,altman15,deluca13,gramsch12,deluca14,huse13,Rahul15,rademaker16,chandran15,ros15,
 eisert15,serbyn141,Rahul15,zakrzewski16,prelovsek2018}  and  logarithmic growth of the entanglement entropy \cite{Znidaric08,bardarson12,kjall14,Serbyn13,serbyn15,luitz16}.
 It has also been found that MBL prevents a driven system from heating \cite{kozarzewski16,abanin2017,abanin2015,Ponte2015,Bordia2017, Choi2017,Zhang2017}.
 These unusual properties  can be  explained via  the existence of a macroscopic number of local integrals of motion \cite{Serbyn2013, huse14,imbrie2017,chandran15,rademaker16,obrien16,inglis16,gluza2017,mierzejewski2018}.

While most of theoretical studies so far concentrated on the one-dimensional (1D) disordered model of interacting spinless fermions,
the experiments on MBL are performed  on cold--fermion lattices \cite{schreiber15,choi16,bordia16,bordia2017_1} where the relevant model is
the Hubbard model with spin--$1/2$ fermions, whereby the disorder enters only via a random (or quasi-periodic) charge potential. 
Recent  numerical studies of such a  model  \cite{mondaini15,prelovsek216,bonca2017,bonca2017,mierzejewski2018} 
reveal that even at strong disorder, localization and nonergodicity occurs only in the charge subsystem, implying a partial MBL. 
Unless one introduces also an additional random magnetic field \cite{gal2017,mierzejewski2018}, the spin remain delocalized,
\cite{potter16,su21,su23,friedman2017,prelovsek216,parameswaran2017}, although the spin transport is anomalously slow and subdiffusive  
\cite{prelovsek216}.

In the present work we focus on the explanation of the slow spin dynamics and subdiffusion within the disordered 1D Hubbard model.
We first demonstrate that in the case of potential disorder and low particle density  the spin dynamics  can be described by a squeezed 
isotropic Heisenberg model, whereby the distribution of the random exchange interactions is singular. Such an effective model
can be studied  numerically quite in detail, but also analytically taking into account that the 1D spin dynamics is dominated by weak links.   
In this manner we show that spin excitations spread over distance $M$ in a characteristic time $t$ such  that $M \propto t^{\alpha}$ with ${\alpha  \simeq \lambda /(d+\lambda)}$, where 
$d$ is the average distance between singly--occupied sites and $\lambda$ is determined by the Anderson localization length  in the noninteracting system. 
While the mapping on the Heisenberg model is valid for dilute systems $d \gg 1$, numerical results for strongly disordered Hubbard model
reveal that the same qualitative spin dynamics remains valid for all densities even up to half--filling. 

{\it Model.--}
Our aim is to establish the spin dynamics in  the disordered Hubbard chain,
\begin{eqnarray}
H&=& -t_{\rm h} \sum_{i \sigma}   (c^{\dagger}_{i+1\sigma} c_{i\sigma}+{\rm H.c.})+\sum_{i} \varepsilon_i  n_i \nonumber \\ 
   && + U \sum_i  n_{i\uparrow} n_{i\downarrow}. \label{hh}
 \end{eqnarray} 
where $n_i=n_{i \uparrow} +n_{i\downarrow} $ and  $n_{i \sigma}=c^{\dagger}_{i\sigma} c_{i\sigma}$. We study the model with $L$ sites
and $N$ electrons, fixing also total spin projection $S^z_{\rm tot}=0$. We assume  a uniform distribution of random charge potentials, $\varepsilon_i \in [-W,W]$ and  set the hopping integral $t_h=1$.
\begin{figure}[h!]
\centering
\includegraphics[width=\columnwidth]{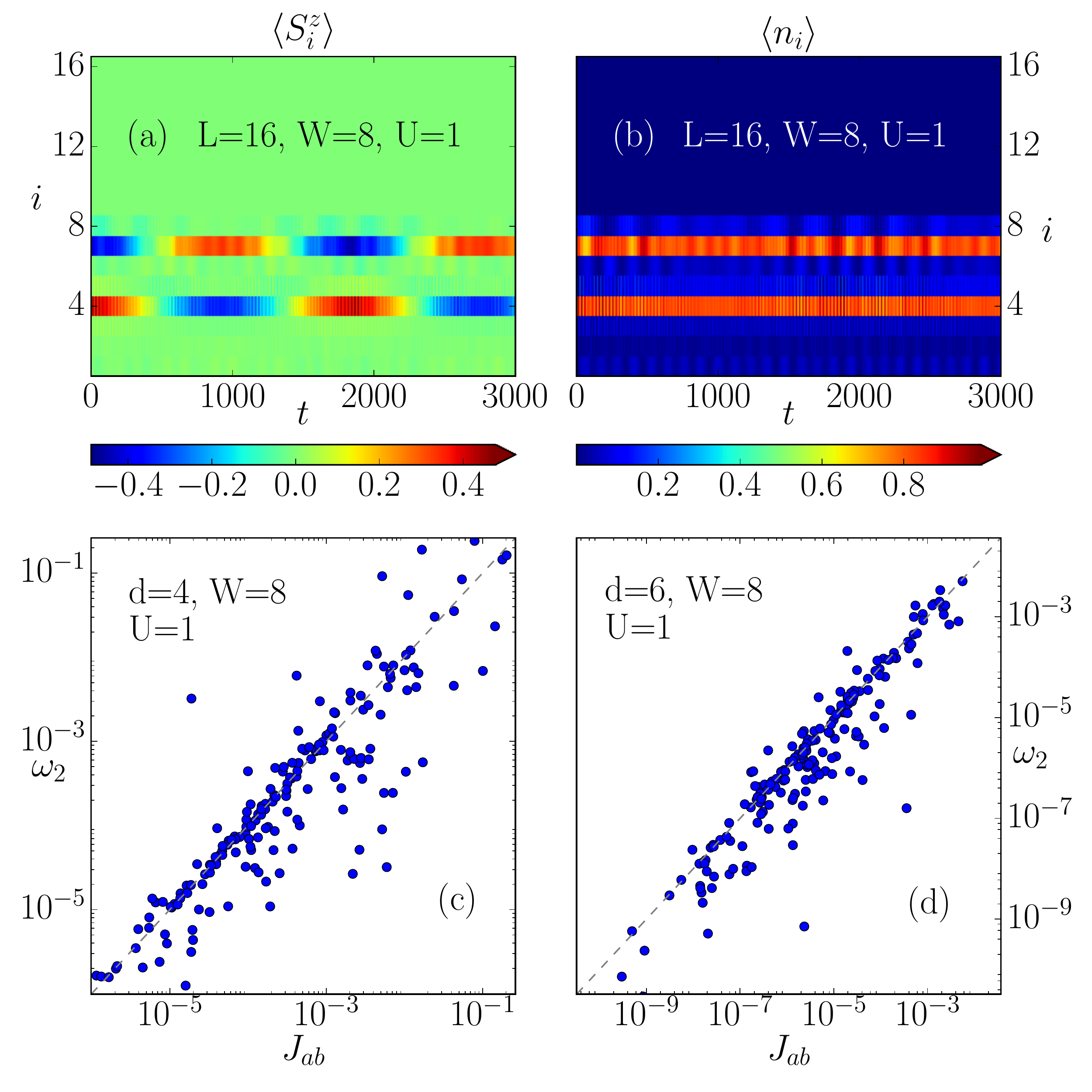}
\caption{ Two electrons on the disordered  Hubbard chain. a) and b):  $\langle S^z_{i} \rangle(t)$  and $ \langle n_{i} \rangle(t)$  for a single 
initial state and single realization of disorder.  c) and d):  frequency $\omega_2$ of spin oscillation obtained directly from the Hubbard model (see panel a) compared with Eq. (\ref{jeff}).  The distance between electrons
is fixed $d=4$ (c) and $d=6$ (d).}
\label{fig1}
\end{figure} 

{\it Two electrons.} In order to gain a preliminary insight to the spin dynamics, we first study two electrons.
The dynamics of a few interacting spinless particles has been studied  previously \cite{shepalyansky1,shepalyansky2,robin2017}. 
Here, we study for illustration $N=2$ electrons with opposite spin projections which  propagate on the chain with $L=16$ sites.
Assuming that particles are initially at sites $j$ and $l$,  $| \psi (0) \rangle = c^\dagger_{j\downarrow} c^\dagger_{j\uparrow} |0\rangle $,   the propagation of  $| \psi (t) \rangle$ is obtained via exact diagonalization.  Figs. \ref{fig1}a and  \ref{fig1}b show, respectively,  time--dependence of the local spin $ \langle S^z_{i}\rangle(t) =  \frac{1}{2} \langle \psi (t) | n_{i \uparrow} - 
n_{i\downarrow}    | \psi (t) \rangle$  and   density  $ \langle n_{i} \rangle(t) =   \langle \psi (t) | n_{i}  | \psi (t) \rangle$
 for one configuration of $\varepsilon_i$ corresponding to  $W=8$.  While for such strong disorder,  the charge degrees appear to be fully
localized,  spins  undergo oscillations. 

{\it Effective spin model.--} The coexistence of almost frozen charges and oscillating spins suggests that one can derive an effective  spin model.
To this end, we use   the Anderson states as the basis, i.e., we use the single--particle eigenstates, $\phi_{i a}=\langle i |a \rangle$, of the noninteracting ($U=0$) model.
Then, 
\begin{eqnarray}
H&=&\sum_{a \sigma }  \epsilon_{a} c^{\dagger}_{a\sigma} c_{a\sigma} +\frac{U}{2} \sum_{a a' b b' \sigma} \chi^{ab}_{a'b'} c^{\dagger}_{a\sigma}  c^{\dagger}_{b \bar{\sigma}}c_{b' \bar{\sigma}}c_{a' \sigma}, \nonumber \\
\chi^{ab}_{a'b'}&=& \sum_i  \phi^*_{i a}  \phi^*_{i b}  \phi_{i b'} \phi_{i a'} . \label{hht} 
\end{eqnarray}
Assuming that the charge-dynamics is frozen, the main effect arising from the presence of the Coulomb interaction comes from terms with either 
$a=a'$, $b=b'$ or $a=b'$, $b=a'$,  and in both cases $a \ne b$.  Then, the Hubbard  term in Eq.~(\ref{hht}) can be written in a SU(2)--invariant form
\begin{equation}
H_U=\frac{1}{2} \sum_{a \ne b }  J_{ab} \left( \frac{1}{4} n_a n_b - \vec{S}_a \cdot  \vec{S}_b \right), \label{hspin}
\end{equation}
where we use standard density and spin operators:  $n_a=n_{a \uparrow}+n_{a \downarrow}$, 
$S^z_a=\frac{1}{2} (n_{a \uparrow}-n_{a \downarrow}) $, $S^+_a= c^{\dagger}_{a\uparrow} c_{a \downarrow} $, $S^-_a=c^{\dagger}_{a \downarrow} c_{a\uparrow}$.
The effective exchange interaction is ferromagnetic
\begin{equation}
J_{ab}=2 U \chi^{ab}_{ab}=2 U \sum_i   |\phi_{ia} |^2  |\phi_{ib} |^2.  \label{jeff}
\end{equation}

In order to test the approximation Eq.~(\ref{jeff}), we consider  $N=2$  electrons with opposite spins, located 
at sites $j$ and $j+d$. We evaluate the spin--oscillation frequency, $\omega_2$, directly from results for the Hubbard model, see Fig. \ref{fig1}a. 
For the same set of $ \varepsilon_i$ we then identify Anderson states $a, b$ which maximize $|\phi_{ja}|^2$  and  $|\phi_{j+d\;b}|^2$,  respectively.
This enables the calculation of $J_{ab}$ from Eq.~(\ref{jeff}) that should lead to  spin oscillations $\langle S^z_{j,l} \rangle = \pm \frac{1}{2} \cos(J_{ab} t )$.
Figs. \ref{fig1}c and  \ref{fig1}d  show correlations between $\omega_2$ and $J_{ab}$ for various realizations of disorder and various distances $d$ between the electrons.
One finds that indeed $\omega_2 \simeq J_{ab}$ for strong disorder $W \gg 1$ and $d \gg 1$.

For low density of carriers and larger disorder,  the maxima of the occupied Anderson states $a$ and $b$ are typically 
separated  by  $x_{ab} > \xi $, exceeding the single-particle localization length, $\xi$. Then, one  obtains an approximate relation
\begin{equation}
J_{ab} \simeq 2U \exp(-x_{ab}/\lambda), \quad  \quad   \lambda \sim \xi \label{jeff1}
\end{equation}

{\it The squeezed spin model.--} Assuming that  charges are frozen to the initial occupations $n_i= 0,1,2$ it is evident that the effective spin model 
Eq.~(\ref{hspin}) acts only on singly occupied sites with $n_i=1$. Spin dynamics of the Hubbard model at high temperatures $T \gg W,U$ can be then studied 
by first  randomly positioning  $N$ electrons on $L$ sites. This allows us to establish the distribution of distances between the  singly occupied sites as well as the distribution 
of the effective $J_{ab}$, using Eq. (\ref{jeff}) or its simplified version, Eq. (\ref{jeff1}).  Due to the exponential decay of $J_{ab}$  we  consider only 
couplings between the neighboring singly occupied sites.  The effective Heisenberg model on a squeezed chain then reads
\begin{equation}
H_H=-\sum_i  J_{i}  \vec{S}_i \cdot  \vec{S}_{i+1}, \label{heff}
\end{equation} 
where the summation is carried out over  singly occupied sites $n_i=1$, $i \in \{1,...,\tilde N\}$ with $\tilde N \leq N$.  
Note that  at infinite temperature, the  average lattice--spacing in the effective model equals $d= L/ \tilde N =(\bar{n}-\bar{n}^2/2)^{-1}$, where $\bar{n}=N/L$ is the average filling in the original Hubbard model.
  
 In order to establish the probability distribution of $J_{i}$, we first consider a section of length $L \gg 1$, 
where we randomly choose the continuous positions of $ \tilde N=L/d $ points and study the regime $L \gg 1$.  The probability density 
for the distances between the neighboring points is 
$f_d(x)=\frac{1}{d}\exp (-x/d)$.  While the latter result has formally been obtained for continuous positions of points,  it should hold true also for discrete positions of singly occupied sites  provided that $d \gg 1 $. Using this result one may find the probability density for the random exchange interaction $f_{J}(J)$. 
To this end, we use Eq. (\ref{jeff1}) and  compare the cumulative distribution functions
\begin{equation}
\int_0^y {\mathrm d}x  \frac{1}{d} \exp \left(- \frac{x}{d} \right) =\int_{2U \exp(-y/\lambda)}^{2U} {\mathrm d} J\; f_J(J). \label{cum}
\end{equation}
Taking the derivative of  Eq. (\ref{cum}) with respect to $y$ and introducing  the dimensionless interaction $\tilde{J}=\frac{J}{2U}$ one gets
\begin{equation}
f_{\tilde{J}} (\tilde{J})=\tilde \lambda  \tilde{J}^{\tilde \lambda-1}, \qquad  \tilde \lambda = \lambda/d. \label{distri}
\end{equation}
It is clear  that the interaction $U$  sets the energy scale (and the time scale)  of the effective model, 
whereas the qualitative spin dynamics depends on the ratio between effective localization length, $\lambda$, and inter--particle distance $d$. 
The important message is that for low doping ($d\gg 1$) and strong disorder ($\lambda \sim 1$)  one obtains $\tilde \lambda \ll 1$ with
 the distribution of $\tilde{J}$ being singular at $\tilde{J}=0$. Still, $\lim_{\delta \rightarrow 0^+}  \int_0^{\delta}   {\mathrm d} \tilde{J}\; f_{\tilde{J}} (\tilde{J})=0$, 
hence the probability for cutting the Heisenberg chain into disconnected subchains is vanishingly small. 

\begin{figure}[h!]
\centering
\includegraphics[width=\columnwidth]{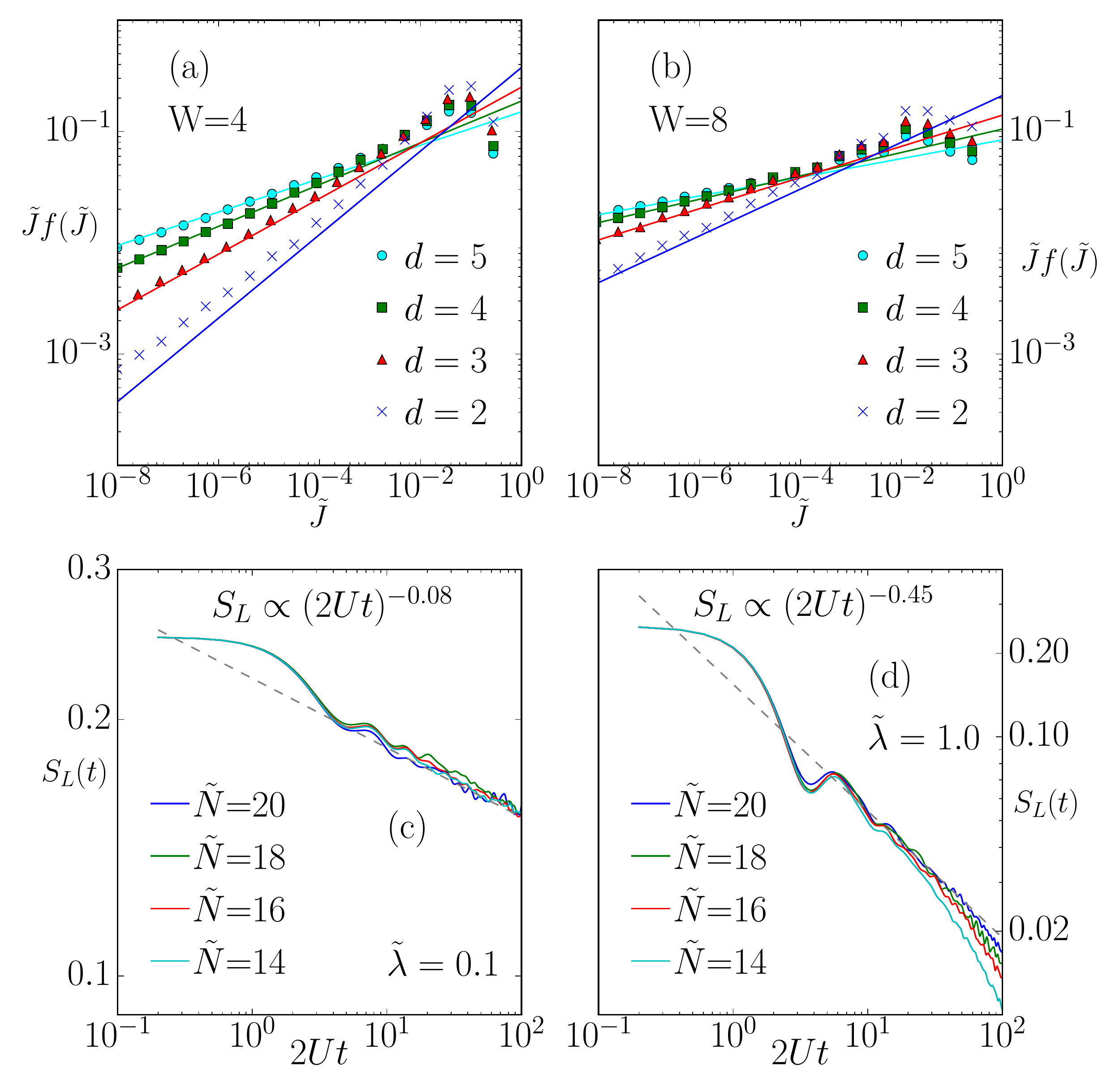}
\caption{Points in a) and b) show $\tilde{J} f_{\tilde{J}} (\tilde{J}) $, generated directly from Eq. (\ref{jeff}) for $N=2$ electrons at average distance $d$,
whereby results have been fitted to Eq. (\ref{distri}) by adjusting a single $\lambda$ (for all $d$). c) and d) local spin correlation  function [Eq. (\ref{spincor})]
for the effective model with various numbers of spins $\tilde{N}$. Results for  $t\in [10,50]$ with largest $\tilde{N}$  are fitted by  $S_{L}(t) \propto t^{-\alpha}$  shown as dashed line.}
\label{fig2}
\end{figure} 

\begin{figure}[h!]
\centering
\includegraphics[width=\columnwidth]{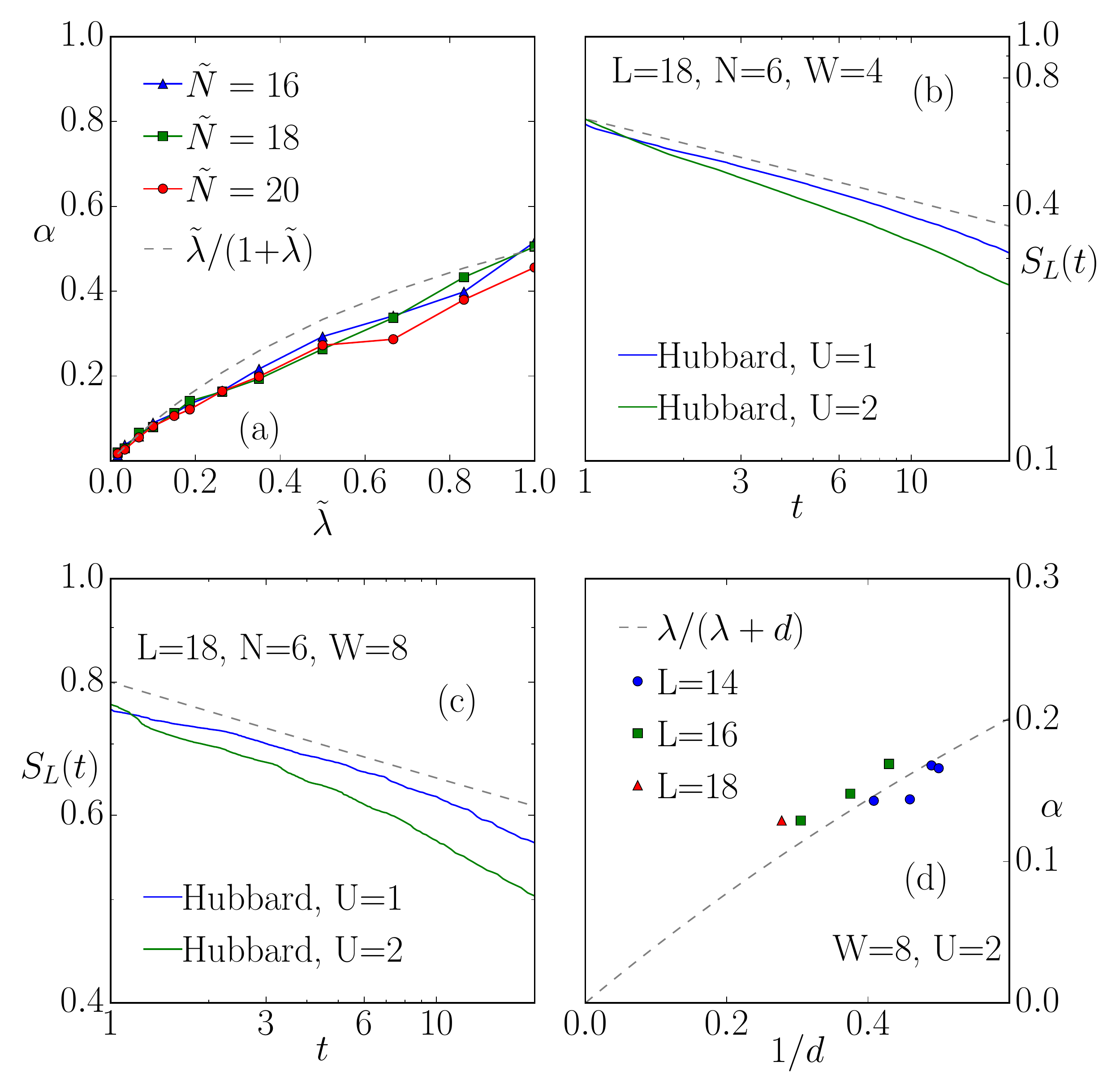}
\caption{a):   Dynamical exponent $\alpha$ vs. $\tilde{\lambda}$ obtained for squeezed model. 
 b) and c): Spin--spin correlation function obtained for the Hubbard model ($L=18$, $N=6$) and compared with $t^{-\alpha}$ (dashed line) where $\alpha=\lambda/(\lambda+d)$.
 d):   $\alpha$ in the Hubbard model ($U=2$, $W=8$, various $N$ and $L$).  $\lambda$ in b)-d) is obtained from fits in Fig. \ref{fig2}b.
}
\label{fig3}
\end{figure} 

In order to test feasibility and accuracy of Eq. (\ref{distri}) we have numerically generated the distribution of 
$\tilde{J}= J_{ab}/2U$ also  directly from  Eq.~(\ref{jeff}),  in the same way as discussed for $N=2$ case.
The positions of two electrons $l$  and $j$ have been randomly chosen in such a way that the distance $x=| l-j | $ 
is distributed according to  $f_d(x) $ for various $d$.  Numerical results for $W=4$ and 8 are shown in 
Figs. \ref{fig2}a and  \ref{fig2}b, respectively. These results have been fitted by Eq.(\ref{distri}),  whereby we adjusted a single parameter
$\lambda$ for all values of $d$.  We have obtained $\lambda \simeq 0.75 $ and  $\lambda \simeq 0.4 $  for $W=4$ and $W=8$, respectively. 
Although Eq. (\ref{distri}) has been derived for $d \gg 1$,  it turns out to remain qualitatively valid also for $d=2$, i.e. for the
average distance between singly--occupied sites in the half--filled Hubbard model.
We conclude that Eq. (\ref{distri}) accurately describes
$f_{\tilde{J}} (\tilde{J})$  at least for small $\tilde{J}$, i.e.,  in the regime which is essential for the long--time spin dynamics.  

{\it Local spin correlations.--} 
We first calculate the time--dependent local spin correlations at infinite temperature,
\begin{equation}
S_L(t)= \langle S^z_i S^z_i(t) \rangle = \frac{1}{\rm Tr\; 1} \langle {\rm Tr } \left[  S^z_i (t) S^z_i \right] \rangle_{\rm dis}  \label{spincor}
\end{equation} 
where the spin evolution is determined  by the effective $H_H$. We take the random interaction $J_{i}=\tilde{J}$ as given by Eq.  (\ref{distri}), i.e.,  
we express time in units of  $1/2U$. $ \langle ... \rangle_{\rm dis}$ means averaging over various realizations of $J_{i}$ and  we use at least $2000$ disorder samples.
  
Figs. ~\ref{fig2}c and  \ref{fig2}d show $S_L(t)$.   For longer times and  $\tilde{\lambda} < 1$ one observes
power-law decay $S_L(t) \propto (2Ut)^{-\alpha}  $ with $\alpha  < 1/2$, hence the spin dynamics is clearly subdiffusive.  
In Fig. \ref{fig3}a we demonstrate $\alpha$ obtained from fitting numerical results  by $S_{L}(t) \propto t^{-\alpha}$ in the  time--window $t \in [10,50]$. 
The main message coming from these studies is that $\alpha >0 $ for arbitrary nonzero $\tilde{\lambda} > 0$, i.e., for arbitrary nonzero filling.
For $\tilde{\lambda} \ll 1$ we obtain the exponent $\alpha \simeq \tilde{\lambda}$. Still, it should be noted that the distribution Eq.~(\ref{distri}) is singular only for $\tilde \lambda <1$ which should be the regime of subdiffusion. For $\tilde{\lambda} \ll 1$ the finite size effects are negligible (Fig.~\ref{fig2}c) but become significant for larger $\tilde{\lambda}$, Fig.~\ref{fig2}d. 
Nevertheless, for the regime with $\tilde \lambda \simeq 1$ (relevant for larger filling $\bar n \sim 1$ and/or weaker disorder)  our results shown in Fig.~\ref{fig2}d are consistent  with normal spin diffusion with $\alpha =1/2$, which is also expected in the weakly disordered Hubbard model.  Results in  Figs. $\ref{fig2}c$, $\ref{fig2}d$ and $\ref{fig3}a$ support the scenario, that the spin excitations spread subdiffusively  due to the singular distribution of random exchange interactions, Eq. (\ref{distri}).

{\it Single weak--link scenario.--}
To explain the relation of the subdiffusive  dynamics and the singular distribution of $\tilde J_i$,  we consider a single spin excitation and 
estimate the time, $t$,  in which the excitation spreads over $M$ sites in the effective Heisenberg chain. We assume the weak--link scenario, where the long--time dynamics
is govern by rare regions with the smallest $J_i$. Similar approach has been used to describe the subdiffusive transport of spinless particles in the vicinity of the MBL transition \cite{agarwal15,agarwal17,luschen2016,bordia2017_1}. 
Here, we assume that $t \sim 1/(2U \tilde J_m)$, where $\tilde{J}_m$ is the weakest exchange out of $\tilde{J}_{i }$ for $i=1,...,M$.
The probability that each random $\tilde{J}_{i}$ is larger than $\tilde{J}_0$ is
\begin{eqnarray}
\left[ \int_{\tilde{J}_0}^{1} {\rm d}\tilde{J}\; f_{\tilde{J}} (\tilde{J}) \right]^M  
=\int_{\tilde{J}_0}^{1} {\rm d}\tilde{J}_m \;f_m(\tilde{J}_{m}), \label{eq2}
\end{eqnarray} 
where  $f_m(\tilde{J}_{m})$ is the probability density for  the smallest interaction.
 Using Eq. (\ref{distri}) and calculating derivative of Eq. (\ref{eq2}) with respect to  $\tilde{J}_0$ one finds  the distribution function
 $f_m(\tilde{J}_{0})= \tilde{\lambda} M  \tilde{J}_{0}^{\tilde{\lambda}-1}  (1-\tilde{J}^{\tilde{\lambda}}_0 )^{M-1} $.
Then, the expectation value of the smallest exchange interaction out of $M$  random $\tilde{J}_i$ reads
 \begin{eqnarray}
 \langle \tilde{J}_{m} \rangle  &= & \int_{0}^{1} {\rm d}\tilde{J}_m \;f_m(\tilde{J}_{m}) \;\tilde{J}_m 
  \simeq   \Gamma \left(1+ \frac{1}{\tilde{\lambda}} \right) M^{-\frac{1}{\tilde{\lambda}}}. \nonumber \\
 \end{eqnarray} 
In the latter equation we have used  
formulas for asymptotics  at $M \gg 1/\tilde{\lambda} $. 
So we find the relation between the spread of the spin excitations $\Lambda$ and $t$ as,
\begin{equation}
\Lambda \sim M d \propto \left( 2U  t   \right)^{\tilde{\lambda}}, \qquad S_L(t) \propto \Lambda^{-1} \propto \left( 2U  t  \right)^{-\tilde{\lambda}}.\label{sdif}
\end{equation}
 The exponent $\alpha=\tilde{\lambda}$  is the same as previously obtained from numerical studies of the effective Heisenberg model for $ \tilde{\lambda} \ll 1$. 

{\it Multiple weak--link scenario.--}  The single--weak link scenario breaks down for $\tilde{\lambda} \sim 1$, where $\alpha \simeq \tilde{\lambda}/2$ instead of  $\tilde{\lambda}$, as shown 
in  Fig.~\ref{fig2}d.
As an alternative explanation for the subdiffusive transport we consider 
distribution of effective {\it hopping times} between neighboring sites in the squeezed spin model. The relevant dimensionless quantity is  $\tau=1/\tilde{J}$. Using Eq. (\ref{distri}) one finds
the probability density $f_{\tau}(\tau)=\tilde{\lambda}/\tau^{\tilde{\lambda}+1}$.  For such broad distribution of hopping times in a {\it classical} model of random traps \cite{bouchaud,machta85}, 
 one gets subdiffusive transport $\Lambda \propto \left( 2U  t   \right)^{\alpha}$ where $\alpha=\tilde{\lambda}/(1+\tilde{\lambda})$ for $\tilde{\lambda}  < 1$.  In the latter model, a classical particle
may hop between neighboring traps in time $\tau$. The hopping time, $\tau$, is randomly chosen for each site but remains the same for each visit of the same site. This simple model quite accurately reproduces the dynamical exponent  $\alpha$ for arbitrary $\tilde{\lambda}$, as shown in Fig. \ref{fig3}a, whereas for $\tilde{\lambda} \ll 1$ it gives the same relation as
the single weak--link scenario.

{\it Comparison with the Hubbard model.--} 
Finally,  we compare our analytical predictions with numerical results obtained directly for the Hubbard model. 
The time-dependent local spin correlation function $S_L(t)$ at infinite temperature has been obtained using the microcanonical Lanczos method (MCLM) \cite{long03,
prelovsek13} (in analogy to the {\it imbalance} correlations presented previously \cite{prelovsek216}) for the Hubbard model with $L=18$ and $N=6$, i.e. $d \simeq 3.5$.
Results are shown in Figs.~\ref{fig3}b, ~\ref{fig3}c  together with analytical prediction, $ S_L(t) \propto (2Ut)^{-\alpha}$ with $\alpha=\lambda/(\lambda+d)$ and $\lambda$ obtained from fits in Fig.~\ref{fig2}b.  
Despite significant finite--size effects, one observes  that the latter estimate  correctly describes the subdiffusive spin dynamics in the Hubbard model at low filling $\bar n \ll 1$. 
In particular, the exponent $\alpha$ obtained directly from  the Hubbard model weakly depends on $U$.   

Moreover, one can consider the validity of the subdiffusion scenario in the full Hubbard model beyond the limit of low filling.
It has previously been  found  that spins reveal a subdiffusive dynamics even for $\bar n =1 $ \cite{prelovsek216}. We therefore analyze the MCLM  results for 
$S_L(t) \propto  t^{-\alpha}$  considering various system sizes $L=14,16,18$ and various numbers of electrons $N$. 
For the time--window $t \in [1,10]$  we extract  the dynamical exponent $\alpha$  and compare with  $\alpha=\lambda/(\lambda+d)$, as shown in Fig. \ref{fig3}d.
Our approach works even up to  $\bar{n}=1$ since the average distance between singly occupied sites  $d \ge  2$,
while $\lambda<1$ provided that the disorder is sufficiently strong. In particular, for $W=8$  and   $\bar{n} \le 1$ we have estimated  that $\tilde{\lambda} \lesssim 0.2$.

 {\it Conclusions.--} In this paper we presented an explanation for the anomalous spin dynamics  in 1D Hubbard model with large potential disorder
 in the regime of partial MBL, where the charge dynamics appears to be frozen whereas  spins exhibit ergodic but subdiffusive transport \cite{prelovsek216}.     
 We have derived an effective isotropic Heisenberg model
with random exchange interactions between neighboring singly--occupied  sites $J_i$.  Our derivation is formally best applicable to the regime of low filling,
$\bar n \ll 1$,   and strong disorder. 
The main origin of the subdiffusive behavior then appears to be the singular distribution of the effective exchange interaction, $J_i$, with the crucial parameter  $\tilde \lambda = \lambda/d$  
representing the ratio of the single-particle Anderson localization length
and the  average distance between singly occupied sites.  Results for the Hubbard model reveal that such scenario seems to remain qualitatively  
valid beyond the considered limits  of low filling,  even at $\bar n=1$, provided that the disorder is sufficiently strong.  
It appears that there is no  threshold filling $\bar n_c$, below which also spins  would become localized and full MBL would prevail.

There are still questions concerning the  dynamics within disordered Hubbard model, being relevant also to cold-atom experiments on MBL \cite{schreiber15,choi16,bordia16,bordia2017_1}.
Our derivation of the effective model, remains on the level of spin dynamics, while  charge degrees are assumed to be frozen. It is evident that higher order terms in the Anderson basis, following from Eq.~(\ref{hht}), would lead also to the dynamical coupling between charge and spin degrees of freedom. Since the spin dynamics is ergodic, it is not excluded that also charges would eventually delocalize, but then on much larger time scales.

\acknowledgments  
  This work is supported by the National Science Centre, Poland via project 2016/23/B/ST3/00647 (MM and MK) and P.P. acknowledges the support by the program P1-0044 of the
Slovenian Research Agency.

\bibliography{references.bib}

\end{document}